\documentclass{article}

\usepackage{graphicx}

\newcommand{\pcc}{\mbox{\,cm}^{-3}}
\newcommand{\psc}{\mbox{\,cm}^{-2}}
\newcommand{\VS}{V\'azquez-Semadeni}
\def\ltsima{$\; \buildrel < \over \sim \;$}    
\def\lesssim{\lower.5ex\hbox{\ltsima}}           
\def\gtsima{$\; \buildrel > \over \sim \;$}    
\def\gtrsim{\lower.5ex\hbox{\gtsima}}           

\title{Diffuse interstellar medium and the formation of molecular clouds}

\begin{document}


\author{
  P. Hennebelle, Ecole Normale Sup\'erieure and Observatoire de Paris \and
 M.-M. Mac Low, American Museum of Natural History, New York \and
E. V\'azquez-Semadeni, Universidad Nacion\'al Aut\'onoma de M\'exico,
  Morelia}

\date{ }

\maketitle


\section{Summary}

The formation of molecular clouds (MCs) from the diffuse interstellar
gas appears to be a necessary step for star formation, as
young stars invariably occur within them. However, the mechanisms
controlling the formation of MCs remain controversial.  In
this contribution, we focus on their formation in compressive flows
driven by interstellar turbulence and large-scale gravitational
instability.

Turbulent compression driven by supernovae appears insufficient to
explain the bulk of cloud and star formation.  Rather, gravity must be
important at all scales, driving the compressive flows that form both
clouds and cores. Cooling and thermal instability allow the formation
of dense gas out of moderate, transonic compressions in the warm
diffuse gas, and drive turbulence into the dense clouds. MCs may be
produced by an overshoot beyond the thermal-pressure equilibrium
between the cold and warm phases of atomic gas, caused by
  some combination of 
the ram pressure of compression 
and
the self-gravity of the compressed gas. 

In this case, properties of the clouds 
such as their mass, mass-to-magnetic flux ratio, and total kinetic and
gravitational energies are in general time-variable quantities. MCs may
never enter a quasi-equilibrium or virial equilibrium state but rather
continuously collapse to stars. 
Gravitationally collapsing clouds
exhibit a pseudo-virial energy balance $|E_{\rm grav}|
\sim 2 E_{\rm kin}$, which however is representative of
contraction rather than of virial equilibrium in this case.
However, compression-driven cloud and core formation 
still involves significant delays as additional material accretes,
leading to lifetimes longer than the free-fall time.
In this case, the 
star formation efficiency (SFE) may be determined by
the combined effect of the dispersive action of the early stellar
products formed in the density fluctuations produced by the initial
turbulence, and of magnetic support of large fractions of the volume of
the MCs.

\section{Large scale Interstellar Medium} \label{sec:large_scale}

At the scale of galactic disks, gravitational instability occurs
not just in the gas alone (Goldreich \& Lynden-Bell 1965), but in the
combined medium of collisionless stars and collisional gas (Gammie
1992, Rafikov 2001).  The combination is always more unstable than
either component in isolation (Gammie 1992). In relatively gas-poor
galaxies like the modern Milky Way, the instability of the stars and
gas is often analyzed independently, in terms of gas flowing into
stellar spiral arms (e.g. Roberts 1969).  However, this is just an
approximation to the general gravitational instability.  Analysis of the
instability must include both mass distributions (Yang et al.\ 2007).

\begin{figure}
\includegraphics[width=9cm,angle=0]{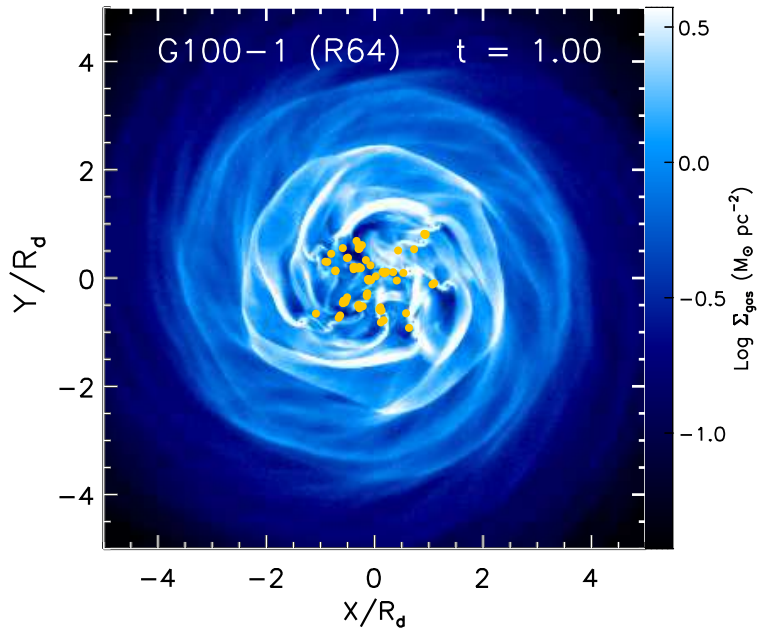}
\caption{Gas surface density map from the models of Li et al.\
  (2005, Fig 5c), showing a model with 6.4 million particles distributed
  evenly between gas and collisionless populations (stars and dark
  matter).  Sink particles are shown in yellow.  The model galaxy has
  rotation velocity at the virial radius of 100~km~s$^{-1}$, and sound
  speed $c_s = 6$~km~s$^{-1}$.}
\label{optional-image}
\end{figure}

Numerical experiments on the behavior of the gravitational instability
in disks have been done by Li et al.\ (2005), using isothermal gas,
collisionless stars, and live dark matter halos computed with GADGET
(Springel et al.\ 2001). They controlled the initial gravitational
instability of the disk, and then computed its subsequent behaviour,
as shown in Figure~\ref{optional-image}.  Using sink particles, they
measured the amount of gas that collapsed as a function of time, and
related it to the initial instability, as expressed by the initial
minimum Toomre parameter for stars and gas combined $Q_{sg, {\rm
min}}$.

\begin{figure}
\includegraphics[width=9cm,angle=0]{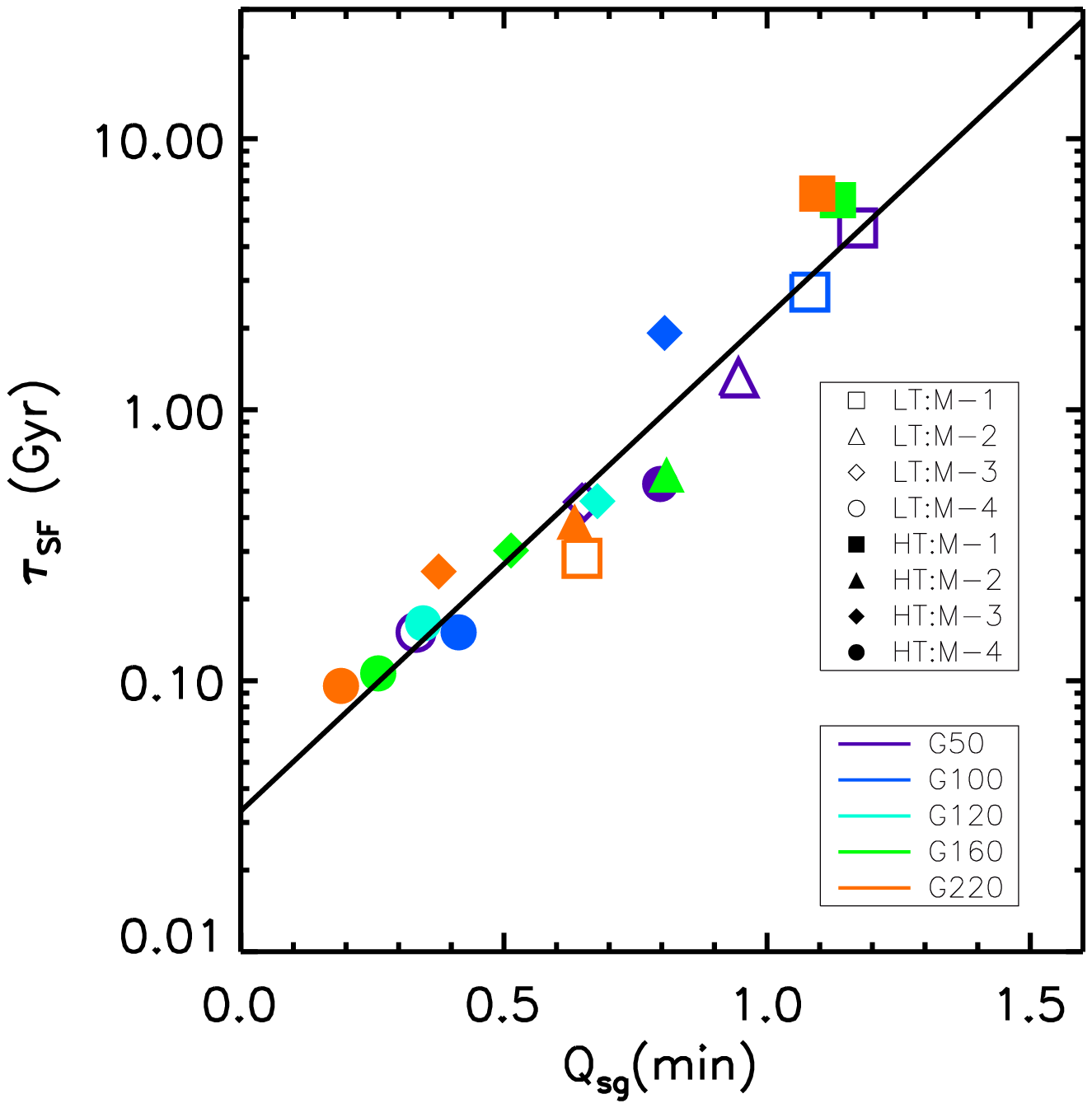}
\caption{Star formation timescale $\tau_{sf}$ correlates exponentially
  with the initial disk instability $Q_{sg, {\rm min}}$ for both
  low-temperature (sound speed $c_s = 6$~km~s$^{-1}$; open symbols)
  and high-temperature ($c_s = 15$~km~s$^{-1}$; filled symbols)
  models. The solid line is a least-squares fit to the data (Li et
  al.\ 2005).}
\label{inst-collapse}
\end{figure}

Li et al.\ (2005) measured the collapse timescale $\tau_{sf}$ by
fitting curves of the form $M_* = M_0 (1 - \exp(-t/\tau_{sf})$, where
the amount of collapsed mass is $M_*$, the initial gas mass $M_0$, and
the elapsed time $t$. Figure~\ref{inst-collapse} shows that the
collapse timescale depends exponentially on the initial strength of
the instability
\begin{equation}
\tau_{sf} = (34.7 \pm 7 \mbox{ Myr}) \exp (Q_{sg, {\rm min}}/0.24).
\end{equation}

\begin{figure}
\includegraphics[width=6cm,angle=0]{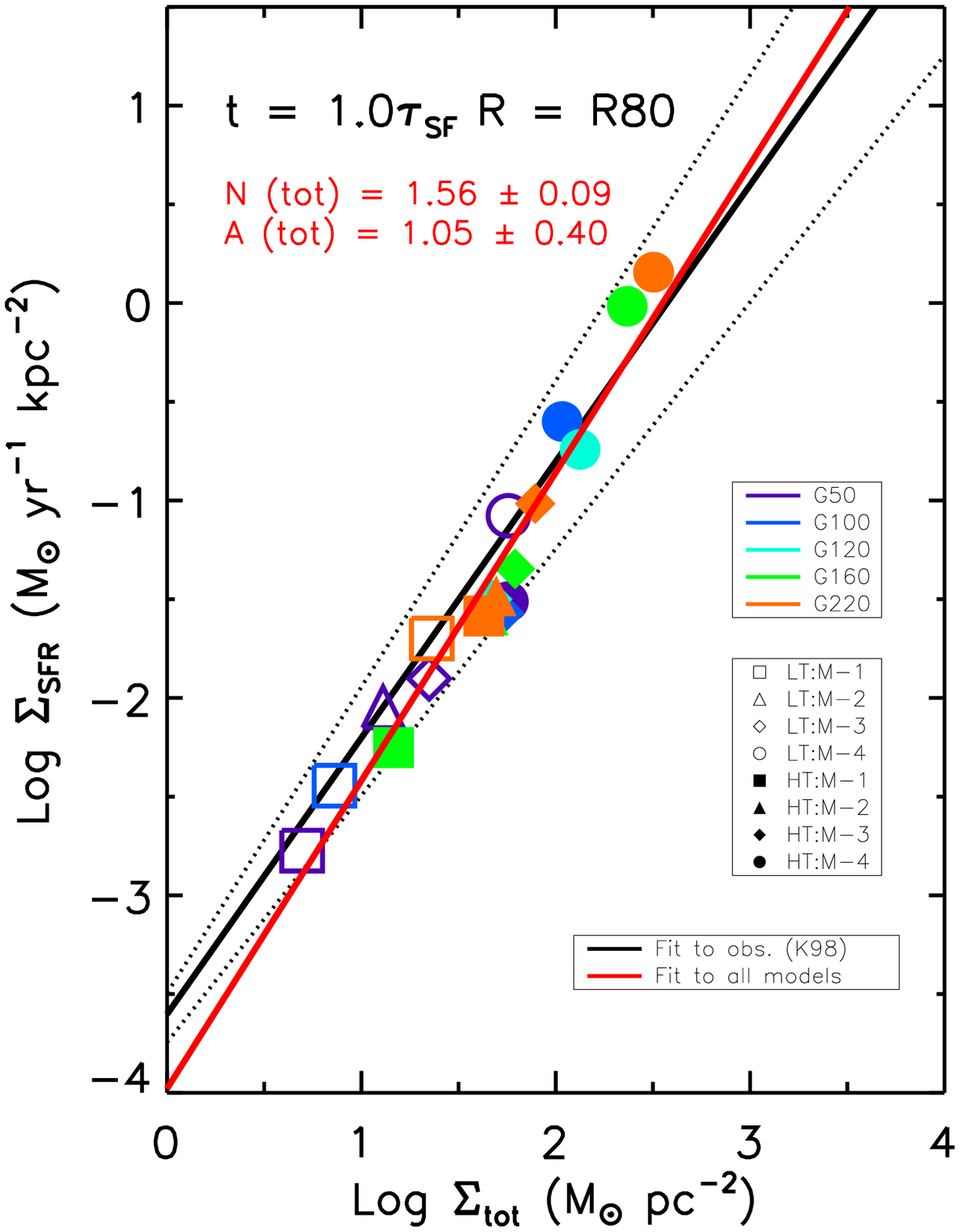}
\caption{Comparison of the global Schmidt laws between the simulations
  of Li et al.\ (2006, Fig. 5) and the observations. The red line is
  the least-squares fit to the total gas of the simulated models, the
  black solid line is the best fit of observations from Kennicutt
  (1998), and the black dotted lines indicate the observational
  uncertainty. }
\label{schmidt}
\end{figure}

Kravtsov (2003) and Li et al.\ (2006) demonstrate that
gravitational instability can explain the Schmidt law (Kennicutt 1998)
as a natural outcome of the evolution of galactic disks. Kravtsov
(2003) computed a cosmological volume and followed the star formation
in individual disks, using a star formation law $\dot{M}_* \propto
\rho_g$ deliberately chosen to not automatically reproduce the Schmidt
law, as compared to the frequently chosen $\dot{M}_* \propto
\rho_g^{1.5}$. The sink particles used by Li et al.\ (2006)
effectively give a similar star formation law, as they measure
collapsed gas above a fixed threshold.  Kravtsov (2003) found that
including feedback made little difference so long as cooling was
prevented below $10^4$~K, roughly the temperature chosen for their
isothermal equation of state by Li et al.\ (2006).  In
Figure~\ref{schmidt} the resulting Schmidt law is shown.

Gravitational instability drives compressive flows at the largest
scales.  Another candidate to drive compressive flows is turbulence
driven by the expansion of H~{\sc ii} regions or supernovae.
Substantial observational evidence exists for multiple generations of
massive star formation in molecular clouds, apparently triggered by
H~{\sc ii} regions and supernova explosions (see, for example,
Elmegreen \& Palou\v{s} 2007).  However, supersonic driven turbulence
inhibits collapse rather than enhancing it (Mac Low \& Klessen 2004 
and references therein).

\begin{figure}
\includegraphics[width=11cm,angle=0]{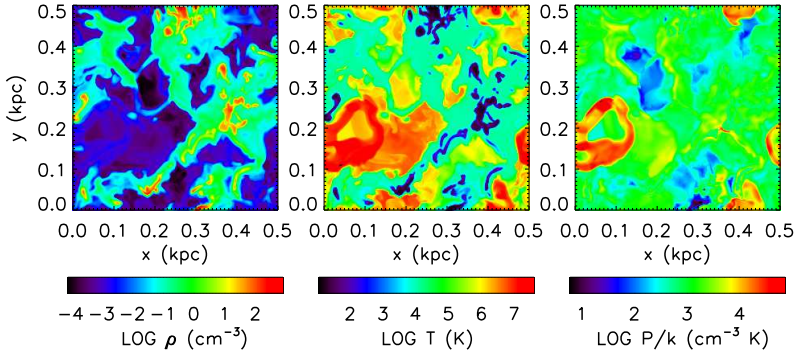}
\caption{Cuts through the midplane (z = 0) of the numerical model of
  Joung \& Mac Low (2006), showing distributions of the density ({\em
  left}), temperature ({\em middle}), and pressure ({\em right}) at $t
  = 79.3$~Myr. Density and temperature both vary by about 7 orders of
  magnitude. Note, however, that high density regions do not correlate
  well with high pressure regions in the absence of self-gravity.}
\label{optional-image-2}
\end{figure}

\begin{figure}
\includegraphics[width=9cm,angle=0]{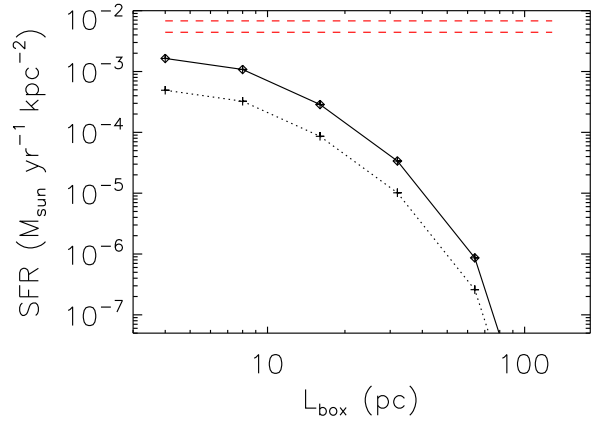}
\caption{Predicted star formation rate from the model plotted against
  the subbox sizes used to measure the Jeans stability of the gas.
  The smallest boxes most accurately measure the amount of
  Jeans-unstable gas available for star formation. (Note that, absent
  self-gravity, collapse doesn't actually occur.)  The dotted line is
  drawn assuming that 30\% of the mass in Jeans-unstable regions turns
  into stars. The dashed lines show the star formation rates
  consistent with the assumed Galactic supernova rate, assuming 130 or
  200 M$_{\odot}$ of stellar mass is required per supernova.}
\label{SFR-jm06}
\end{figure}

An examination of the triggering effect of H~{\sc ii} regions in the
Milky Way by Mizuno et al.\ (2007) led to the conclusion that,
although triggering does occur, only 10--30\% of star formation occurs
in triggered regions. A numerical model, shown in
Figure~\ref{optional-image-2}, of large-scale, supernova-driven
turbulence by Joung \& Mac Low (2006) allowed them to reach a similar
conclusion.  They measured the mass of gas that was Jeans-unstable in
their flow, allowing them to estimate the star formation rate that
would be induced by the turbulent motions alone.  They found that the
corresponding star formation rate was an order of magnitude lower than
the star formation rate required to maintain the assumed supernova
rate (Fig.~\ref{SFR-jm06}).  This again suggests that triggering is a
10\% effect.

Gravitational instability thus appears capable of explaining observed
star formation rates, while turbulent compression alone seems to fail
by roughly an order of magnitude. Considering galactic scale magnetic
fields does not alter this conclusion.  Parker instabilities alone
appear insufficient to form giant molecular clouds (J. Kim et al.\ 2000,
Santill\'an et al.\ 2000), but magneto-Jeans instabilities are
effective (W.-T. Kim \& Ostriker 2006).  

The first consequence of gravitational instability at large scales,
though, is large-scale compressive flows, first in the spiral arms
that form the first manifestation of gravitational instability in
disks, and then in smaller collapsing regions (Field et al. 2006).
Therefore, a general 
physical understanding of molecular cloud formation in compressive
flows is a vital link in understanding star formation.

\section{Neutral Interstellar Medium}

The interplay between compressible turbulence and heating and
cooling processes in the neutral interstellar gas leads to the production of
density fluctuations, including tiny-scale atomic structures 
cold atomic clouds, and ultimately molecular
clouds.
 
Observationally, the neutral atomic hydrogen is mostly studied through
the HI 21cm line, both in emission and in absorption (e.g. Dickey \&
Lockman 1990; Heiles \& Troland 2005).  These observations show that
the interstellar atomic hydrogen spans a wide range of densities and
temperatures, from what has traditionally been called the warm neutral
medium (WNM), with $n \simeq 0.3$--$0.5 \pcc$ and $T \simeq
5000$--$8000$ K, to what has been called the cold neutral medium
(CNM), roughly $100$ times denser and colder, although a significant
fraction of the gas mass (possibly up to 50\%) lies at intermediate
values of $n$ and $T$ (Heiles 2001). This picture is significantly
more complex than the classical two-phase model of Field et al.\
(1969), which proposed the existence of discrete phases (the WNM and
the CNM) in pressure equilibrium and mediated by contact
discontinuities. In the remainder of this section, we discuss our
present understanding of the extent to which this picture is modified
by the presence of compressive turbulence in the atomic gas.  This
turbulence turns out to be transonic with respect to WNM sound speed
but is supersonic with respect to CNM sound speed.

\subsection{Thermal balance and thermal instability}

The detailed thermal balance of the interstellar medium was first
investigated by Field et al.\ (1969) and Dalgarno \& McCray (1972),
and more recently by Wolfire et al.\ (1995, 2003).  The interstellar
atomic hydrogen is thought to be mainly heated by ultraviolet and soft
X-ray radiation through the photoelectric effect on small dust grains
and polycyclic aromatic hydrocarbons.  The heating term is therefore
proportional to the gas number density, $n$, and has a weak dependence
on temperature.  On the other hand, for temperatures larger than a few
thousand Kelvins, the most important cooling mechanism is due to the
Lyman-$\alpha$ HI line and, at lower temperatures, to the [O$_I$] and
[C$_{II}$] fine-structure lines. Since these lines are excited by
collisions, the cooling term is proportional to $n^2$. Thus the gas
is thermally {\it unstable} when the cooling function varies slowly
with the temperature, as is the case at a few thousand Kelvins, when
the [O$_I$] and [C$_{II}$] lines are saturated, and thermally {\it
stable} when the cooling function varies rapidly with the temperature,
as happens in the wings of the atomic lines responsible for the
cooling. This is the physical origin of the famous two-phase model for
interstellar atomic hydrogen (Field et al.\ 1969).

Field (1965) showed that the criterion applicable to the atomic
interstellar gas is the isobaric stability condition (see, e.g., the
review by \VS\ et al.\ 2003), which can be written as
\begin{equation}
\left( {\partial P \over \partial \rho } \right) _{\cal L} \le 0, 
\end{equation}
where ${\cal L}$ is the loss function equal to the cooling minus
heating terms.  When the pressure along the cooling curve decreases
with increasing density, a density fluctuation tends to be amplified
because the surrounding gas has a larger thermal pressure and
compresses it further.

\begin{figure}
\includegraphics[width=9cm,angle=0]{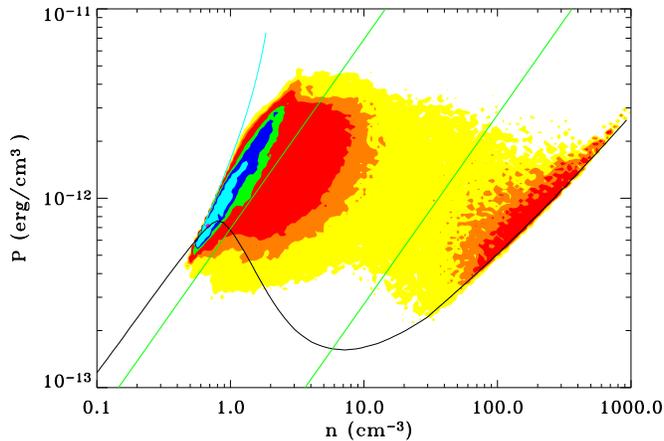}
\caption{ Thermal equilibrium curve for the ISM ({\em black line}),
and isothermal curves with $T=5000$ K and 200 K ({\em green lines}).
The color  represents the gas mass fraction in a simulation of a
turbulent two-phase medium (from Audit \& Hennebelle  2005). 
In arbitrary units, yellow, orange, red, green, dark blue, light blue and pink 
correspond to 1, 5, 10, 50, 100, 200 and 1000.}
\label{curve_therm}
\end{figure}

The cooling curve computed by Wolfire et al.\ (1995, 2003) presents in
a pressure-density diagram, two branches of positive slope joined by
one branch of negative slope. The WNM branch cannot exist at pressures
higher than a certain critical value $P_{\rm max}$, and the CNM branch
cannot exist at pressures lower than another critical value, $P_{\rm
min}$ (see Fig.~\ref{curve_therm}).  The density and temperature of
the stable branches are in good agreement with the density and
temperature ranges of the WNM and CNM inferred from observations.

\subsection{Dynamical formation of cold atomic clouds}

The influence of turbulence on the interstellar gas and its role on
cloud formation was first investigated by V\'azquez-Semadeni et al.\
(1995), who used a cooling curve that implied thermal stability at all
temperatures, although still causing denser gas to be colder. They
found that converging flows form strong density fluctuations even in
relatively weak HII region driven turbulence, because of the strongly
compressible nature of the flow.  Such converging flows have been
proposed to rapidly form molecular clouds (Ballesteros-Paredes et al.\
1999b; Hartmann et al.\ 2001).

The influence of dynamical motions on the warm stable phase of a
thermally bistable flow (i.e. the WNM) was considered in 1D by
Hennebelle \& P\'erault (1999), Koyama \& Inutsuka (2000) and \VS\ et
al. (2006).  These studies showed that either a shock or a converging
flow can trigger the formation of a long-lived cold structure.  The
unperturbed incoming WNM flow undergoes a shock that heats the gas and
throws it out of thermal equilibrium.  Behind the shock, the gas
continues to flow and cool, until finally, roughly one cooling length
behind, it undergoes a transition to the cold phase, forming a thin,
cold, dense layer.  The phase transition occurs provided that the
fluctuation amplitude is sufficiently strong to reach the pressure
threshold $P_{\rm max}$ and that the fluctuation lasts long enough for
the gas to cool.  Perturbations that fail to satisfy either of these
conditions produce weak fluctuations of WNM instead. The response of
the flow is therefore very nonlinear and depends sensitively on
whether the perturbation is able to push the gas into the thermally
unstable area.

The detailed evolution of density and velocity perturbations in a
thermally unstable medium as a function of their associated crossing
times was investigated further  by S\'anchez-Salcedo et al.\ (2002) and \VS\ et
al. (2003) starting with initially thermally unstable gas. 
For {\it density} perturbations whose crossing time is shorter
than the cooling time,  the gas condenses more or less  
isobarically, while if the crossing time exceeds the
cooling time, then the gas follows the thermal equilibrium curve while
condensing. On the 
other hand, {\it velocity} perturbations with crossing times shorter
than the cooling time tend to behave adiabatically, at least before the
gas has time to cool.

In the 1D studies, once the cooling gas reaches the CNM branch, it cools
and contracts along it, 
until the thermal pressure equals the ram pressure of the shocked warm
incoming gas. 
When the perturbation has relaxed, the density within the structure, decreases 
until its internal  pressure is equal to the thermal pressure of the surrounding WNM. 
The cold structure is then pressure confined, and therefore stable, unlike what happens
in an isothermal medium. In 2D and 3D, however, the evolution is more
complicated, as we describe below.

\subsection{Front stability and thermal fragmentation}

By performing a linear stability analysis, Inoue et al.\ (2006) show
that evaporation fronts between cold and warm phases suffer
instabilities similar to the Darrius-Landau instability that occurs in
combustion fronts, rendering them unstable under corrugational
deformations. On the other hand, condensation fronts are stable. The
fastest growth rate of the evaporation front instability corresponds
to wavelengths slightly larger than the Field length, while larger
wavelengths grow proportionally to the wavenumber $k$, and smaller
wavelengths are stable.  Numerical simulations performed by Kritsuk \&
Norman (2002) and Koyama \& Inutsuka (2006) show that the nonlinear
development of the instability can sustain weak turbulence. The source
of energy is the heating term, which is not fully compensated by the
cooling term within the thermal front.

Koyama \& Inutsuka (2002) have also investigated the propagation of a
shock through the WNM in 2D.  They found that the post-shock gas is
very unstable and fragments into many small CNM structures.  Pittard
et al.\ (2005) have shown that radiative shocks become more prone to
overstability at a given upstream Mach number as the final postshock
temperature is lower with respect to the upstream one. All of these
results suggest that weak turbulence can easily be driven in the cold
gas.  Koyama \& Inutsuka (2002) and Heitsch et al.\ (2005) have
further shown that the CNM structures have a velocity dispersion that
is a fraction of the sound speed of the warm phase, but still
supersonic with respect to the internal sound speed of the CNM clumps.

Similar results for a shock-bounded layer in a radiatively cooling gas
have been obtained by Walder \& Folini (1998, 2000). Indeed,
shock-bounded layers were shown to be nonlinearly unstable even in the
isothermal case by Vishniac (1994), through the nonlinear thin-shell
instability.

\subsection{Colliding flows and thermally bistable turbulence}
The formation of CNM structures induced by dynamical motions within
the WNM has been further studied in 2D by Audit \& Hennebelle (2005)
and Heitsch et al.\ (2005, 2006) and in 3D by V\'azquez-Semadeni et
al.\ (2006) (see also Kritsuk \& Norman 2002 for decaying thermally
bistable turbulence and Gazol et al.\ 2005 for the driven case).
Typically, these studies consider a computational box of a few tens of
parsecs and a resolution of about 1000$^2$ cells in 2D, and up to
$400^3$ cells in 3D.  They all consider a converging flow of WNM that
produces a turbulent shocked layer, which fragments into CNM
structures, producing a
turbulent, clumpy medium. 

When the incoming flow is initially nearly laminar, the turbulence in
the dense layer is believed to be driven by the combined action of the
thermal, Kelvin-Helmholz and nonlinear thin-shell instabilities
(Heitsch et al.\ 2005, 2006). However, when nonlinear fluctuations are
initially present, the importance of these instabilities remains to be
clarified (astrophysical flows are not expected to be initially
laminar).  Indeed, Audit \& Hennebelle (2005) find that adding
non-linear perturbations to the converging flow strongly changes the
level of turbulence within the computational box, including in the CNM
phase.

\begin{figure}
\includegraphics[width=8cm,angle=90]{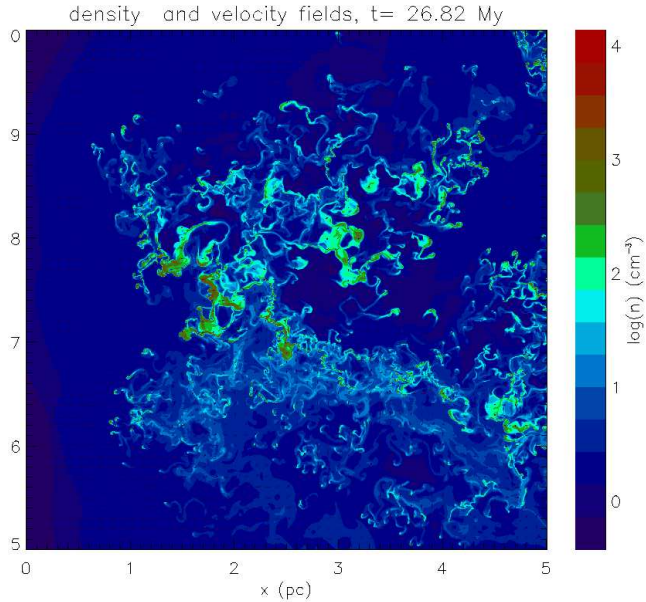}
\caption{Turbulent, two-phase interstellar atomic medium.
Density field of a high-resolution 2D simulation (from Hennebelle \& Audit 2007).}
\label{dens_sim}
\end{figure}

The overall structure of the resulting flow appears to be complex (see
Fig.~\ref{dens_sim}).  The two phases are strongly mixed and a
substantial fraction of thermally unstable gas is produced by the
turbulent motions (Gazol et al.\ 2001) as shown in
Figs.~\ref{curve_therm} and \ref{fig:dens_hist}, in agreement with
observational determinations (Heiles 2001). Note however, that even
when the turbulence is strongly driven, producing transonic velocity
dispersion within WNM, the tendency towards two-phase behaviour
persists, although the fraction of thermally unstable gas increases
for more strongly turbulent regimes. In particular, in the simulations
that have been performed, a large fraction of the CNM structures
remain bounded by contact discontinuities and pressure confined by the
surrounding WNM. Note that it is certainly the case, that for high
Mach number conditions, this structure may change. The question of how
and at which Mach number, remains an open issue.

As in the  numerical experiment of Koyama \& Inutsuka (2002), the CNM structures
present velocity dispersion somewhat smaller than the WNM sound speed. Since these 
motions are supersonic with respect to the CNM internal sound speed, the CNM structures
undergo high-Mach number collisions, which create strong density fluctuations maybe reminiscent 
of the tiny small atomic structures observed in HI (Heiles
1997). 
Note that interestingly enough, this relatively high velocity dispersion within 
CNM structures is  sustained self-consistently by the WNM turbulence, which in the study 
of colliding flows is triggered from the turbulence in the incoming WNM 
as well as by the turbulence spontaneously generated as part of the cloud formation process.

\subsection{Dense structure statistics in thermally bistable turbulent flows}
Studying the statistical physical properties of the dense structures 
produced by interstellar turbulence is of great interest since such
structures constitute both atomic and molecular clouds in the ISM
(Sasao 1973, Elmegreen 1993, Ballesteros-Paredes et al.\ 1999a).
The statistics from the numerical models
can be compared with the observational properties of dense clouds such as,
for example, mean densities, pressures and temperatures, typical sizes,
scaling relations among variables, etc., allowing tests of the various 
theoretical models. 

\begin{figure}
\includegraphics[width=8cm,angle=0]{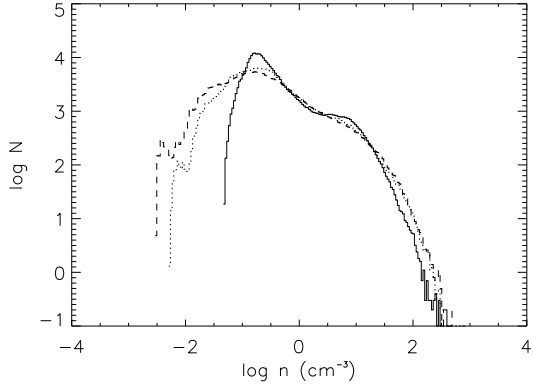}
\caption{Density PDF as a function of rms Mach number $M$ (with respect to
the warm unstable gas). {\it Solid line:} $M=0.5$. {\it Dotted line:}
$M=1$. {\it Dashed line:} $M=1.25$. 
(from Gazol et al.\  2005).
}
\label{fig:dens_hist}
\end{figure}

The simulations by Gazol et al.\ (2005) and Audit \& Hennebelle (2005)
have shown that even at Mach numbers $\sim 1$, local values of the
density and pressure can be reached that exceed $1000 \pcc$ and $\sim
10^5$ K cm$^{-3}$, respectively.  Also, the bimodal nature of the
density histogram, which in the extreme case of a static two-phase
medium would be two Dirac delta functions at the densities of the WNM
and CNM, is seen to gradually loose its bimodal character as the rms
Mach number is increased (Fig. \ref{fig:dens_hist}), signaling the
increasing amounts of thermally unstable gas in the flow. It is also
interesting that sometimes the highest pressures can be obtained in
the transient unstable warm gas, rather than in the densest gas (see
also \VS\ et al.\ 2006).

\begin{figure}
\includegraphics[width=9cm,angle=0]{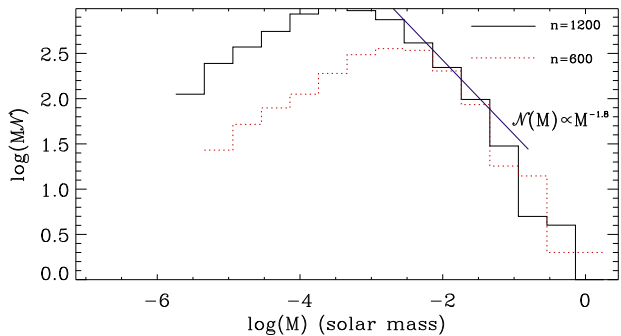}
\caption{Mass distribution of the structure identified in the
  simulations (Audit \& Hennebelle 2008).}
\label{struct}
\end{figure}

\begin{figure}
\begin{center}
\includegraphics[width=6cm,angle=0]{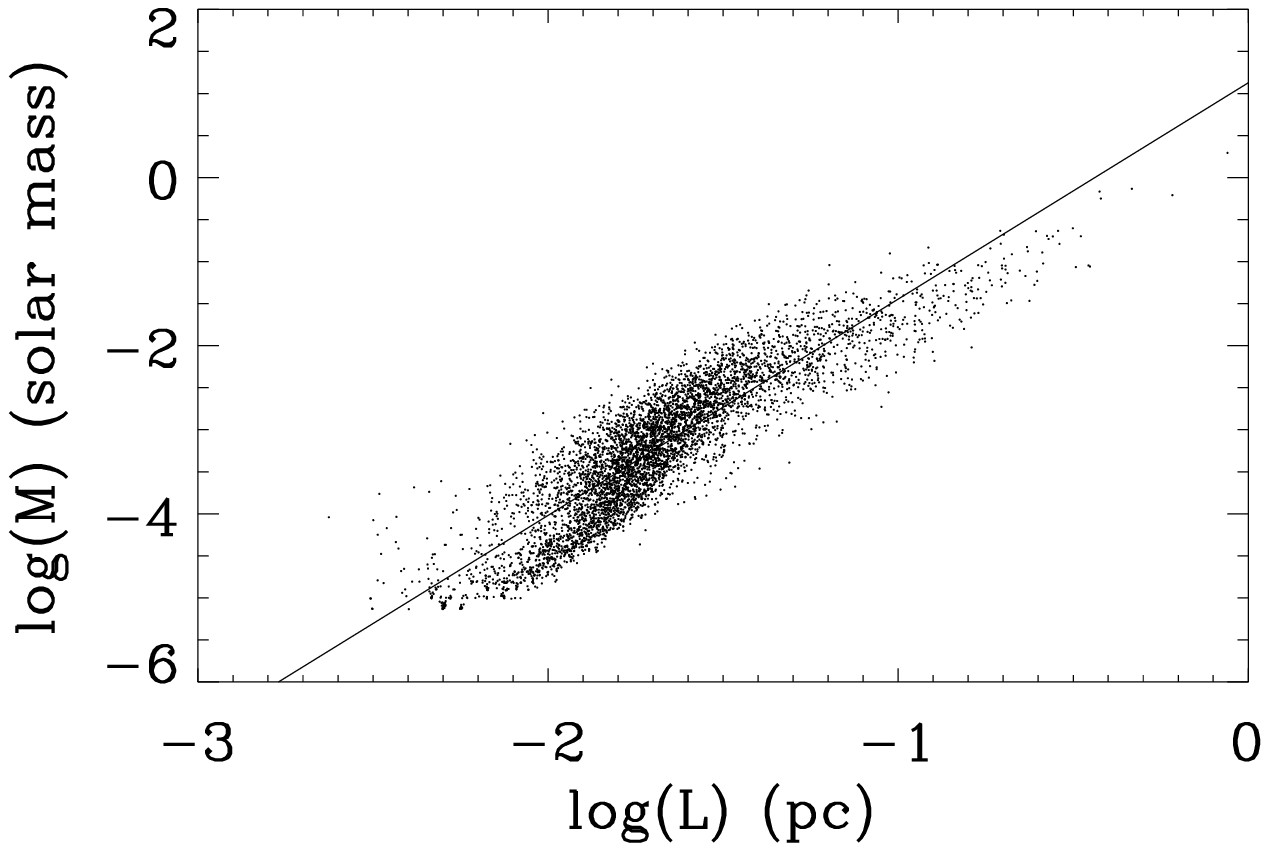}
\includegraphics[width=6cm,angle=0]{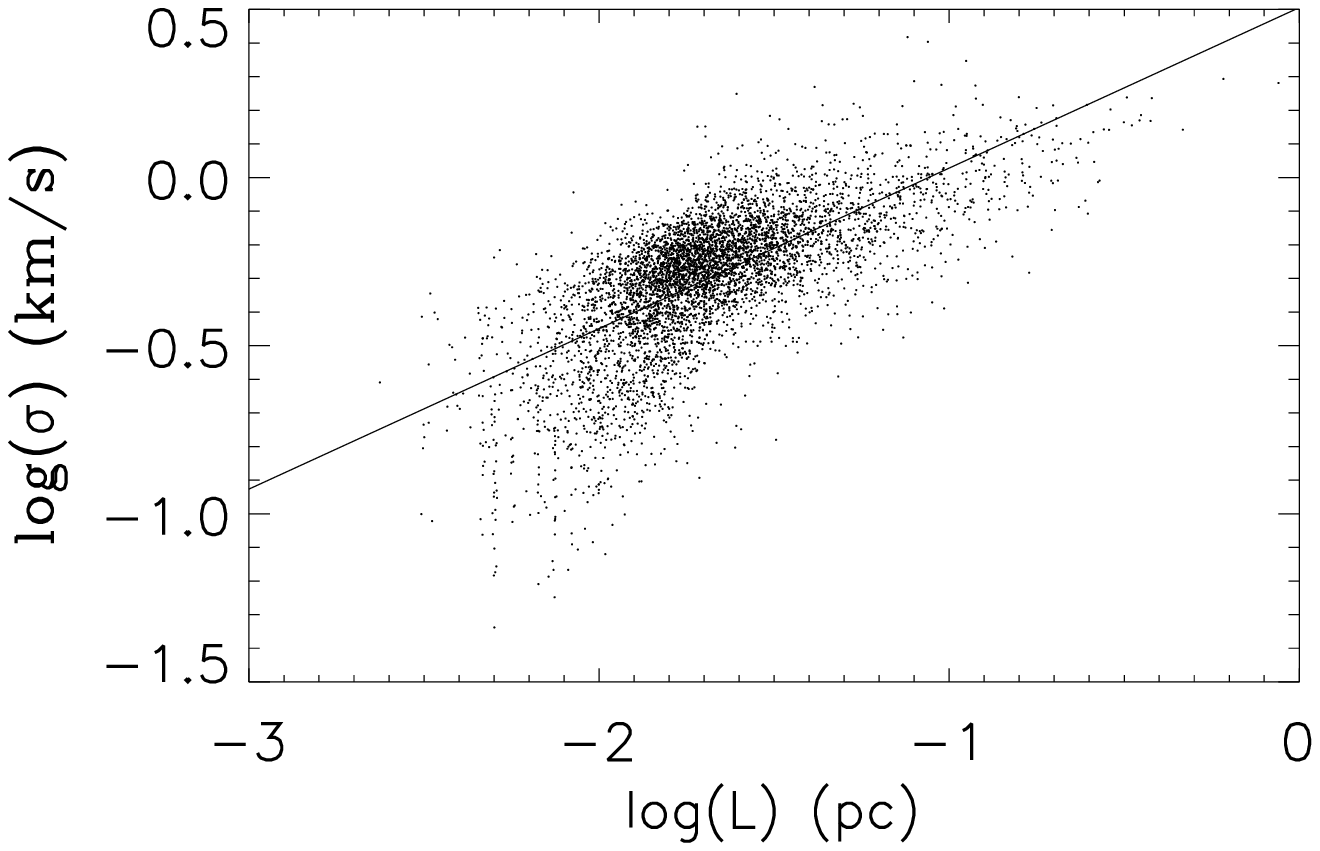}
\end{center}
\caption{Mass versus size ({\em left}) and velocity dispersion versus
  size ({\em right}) relations for the CNM structures extracted from
  the $1200^3$ cells  simulations (Audit \& Hennebelle 2008). The slope of the solid lines are
  2.5 (left panel) and 0.5 (right panel).} 
\label{struct1}
\end{figure}

More recently, much higher resolution simulations have been performed,
using up to $10000^2$ cells in 2D (Hennebelle \& Audit 2007) and
$1200^3$ cells in 3D (Audit \& Hennebelle 2008). Figures~\ref{struct}
and~\ref{struct1} show the mass spectrum, the mass-size relations and
the velocity dispersion as function of size of the CNM structures
extracted from the 3D simulations. Corresponding results for the 2D
case can be found in Hennebelle \& Audit (2007) and Hennebelle et al.\
(2007).  The CNM structures are simply determined by a clipping
algorithm that uses a density threshold lying in the thermally
unstable domain.  The mass spectrum follows $dN/dM = {\cal N} \propto
M^{-1.8}$.  Numerical convergence seems to have been reached at
resolution of $1200$ zones for masses between $\simeq 3 \times
10^{-3}$ and $\simeq 3 \times 10^{-2} M_\odot$.  The mass-size
relation is given approximately by $M \propto L^{2.3-2.5}$, the
smallest value of the index being obtained for the largest
structures. Finally, the internal velocity follows about $\sigma
\propto L^{0.5}$. These indices are very similar to the values
inferred from observations of CO clumps. In particular, Heithausen et
al.\ (1998) (see also Kramer et al.\ 1998), probing clumps of mass as
small as one Jupiter mass, therefore directly comparable with the
masses of the clouds produced in the simulation, report a very similar
mass spectrum and mass-size relation.

It is worth comparing these results with results obtained in supersonic isothermal turbulent
flows. Ballesteros-Paredes \& Mac Low (2002) (see also V\'azquez-Semadeni et al.\ 1997)
have extracted the clumps formed in their simulations both in physical space and 
by a procedure mimicking the observational one, using a clump finding algorithm. 
In both cases,  they found  a lognormal mass spectrum and an internal velocity dispersion
$\sigma \propto L^{0.5}$ compatible with the observations. The mass-size relation appears to be 
more complex. In physical space, they found no correlation between  the mean density 
and size of the clumps. However, structures extracted from the observational procedure followed
approximately $M \propto L^2$. They concluded that this is a projection
effect that artificially 
connects structures along the line of sight. 

It appears therefore that the two-phase model and the supersonic isothermal turbulent 
model lead to statistically different structure distributions. In principle, 
this could constitute a nice test. One should however keep in mind, that at this point, 
the  formation of  molecular hydrogen has not been included in the two-phase model and that structures have 
been extracted in the physical space. Further work is needed before definite conclusions 
can be reached.

\subsection{Influence of the magnetic field}

Hennebelle \& P\'erault (2000) consider in 1D the formation of a single structure when 
the incoming flow makes an angle with the magnetic field.  They find that the thermal condensation 
is possible provided that the angle between the flow and the field is sufficiently small. In this 
process, the magnetic tension plays an important role in unbending the
field lines and therefore
reducing the magnetic pressure. As a result,  the condensation occurs mainly along the field
lines. Thus the magnetic intensity does not increase with the gas density. This seems to 
be in good agreement with the observations (Troland \& Heiles
1986). 
The role of the magnetic waves has been further investigated by Hennebelle \& Passot (2006).
They conclude that over a large range of parameters, the waves tend to trigger the formation of CNM 
structures rather than preventing the condensation.

Recently, the formation of cold structures in 1D via ambipolar
diffusion has been investigated by Inoue et al.\ (2007). They show
that the value of the final magnetic intensity within the structure is
relatively independent of the value of the magnetic field within the
WNM, implying a weak correlation between the density and the magnetic
intensity (see also Heitsch et al.\ 2004).  Such a weak correlation
between the magnetic field strength and the density is a general
feature of MHD turbulence simulations, both isothermal (e.g., Padoan
\& Nordlund 1999; Ostriker et al. 2001) and multi-temperature (e.g.,
Passot et al. 1995; de Avillez \& Breitschwerdt 2005). In the case of
an isothermal gas, the weakness of the correlation has been
interpreted by Passot \& V\'azquez-Semadeni (2003) in terms of the
fact that the various types of (nonlinear, or simple) MHD waves are
characterized by different scalings of the field strength with the
density. Thus, in a turbulent flow in which all kinds of waves pass
through one given point, the field strength there is a function of the
history of wave passages, rather than of the local density.

Thus, it can be seen that in general, a significant correlation
between magnetic field strength and density is {\it not} expected in
the atomic gas as a consequence of the ability of the gas to flow
freely along field lines, except perhaps at the highest densities,
which probably require focused external compressions, which in turn
increase the magnetic pressure as well (Gazol et al.\ 2007). This
seems to be in good agreement with the observations (Troland \& Heiles
1986; Heiles \& Troland 2005).

Concerning the interaction of the magnetic field with the nonlinear
thin shell instability, Heitsch
et al.\ (2007) have begun to investigate the role of the field in
possibly suppressing the instability in the isothermal case, finding
that under a variety of circumstances it may still grow. Studies in the
presence of thermal bistability are still pending.

\section{Formation of molecular clouds} \label{sec:MC_form}

\subsection{Context} \label{sec:context}

MCs are the densest regions in the ISM. Their
distribution in external galaxies shows that giant molecular clouds
(GMCs) are in general the 
     tip of the iceberg
of the gas
distribution, appearing at column densities above roughly 8 $M_\odot
\psc$ (e.g., Blitz et al.\ 2007). As the latter authors conclude, MCs
seem to form out of the HI gas. Their formation thus requires
significant compressions of the diffuse gas. As discussed in \S
\ref{sec:large_scale}, the ultimate drivers of these compressions may be
several large-scale instabilities, such as the global gravitational
instability of the combined stars and gas, the magneto-Jeans instability, the
magneto-rotational instability, or else local disturbances such as
passing supernova shocks, or the general motions of the transonically
turbulent warm diffuse ISM (e.g., Kulkarni \& Heiles 1987; Heiles \&
Troland 2003).

Whatever the source of the compressions, the results described in the
previous sections suggest that transonic compressions in the WNM can
induce a transition to the CNM, followed by an overshoot to physical
conditions typical of GMCs, which are colder and denser than the CNM
clouds. This occurs because the compressed gas is in pressure balance
with the total pressure of the inflowing WNM, including its ram
pressure, implying that the gas at MC densities must be systematically
overpressured with respect to the mean thermal pressure in the ISM
(\VS\ et al.\ 2006), as is indeed the case (e.g., Blitz \& Williams
1999). Moreover, as described in the previous sections, the formation
of dense clouds by this process naturally produces at least some of
the turbulence observed within them.

However, transonic compressions in the WNM alone are not
sufficient to produce the extreme densities and pressures found in the
interiors of MCs. For example, \VS\ et al.\ (2006) found that the
pressure of the dense gas (there defined as gas with $ n > 100 \pcc$)
formed by Mach 2.5 compressions lies in the range
1.5--4 times the mean ISM pressure, while mean observed
MC pressures are actually closer to 10 times the mean ISM
pressure (e.g., Blitz \& Williams 1999). This suggests that
self-gravity must be crucially involved in the formation and evolution of
MCs, to produce the observed additional pressure enhancement.

\subsection{Numerical Results} \label{sec:MCform_results}

The process of dense cloud formation out of compressions in the WNM in
the presence of thermal bistability and self-gravity has been first
studied by \VS\ et al.\ (2007) by means of numerical simulations using
GADGET (Springel et al.\ 2001), which, being Lagrangian, allows very
high effective resolution at very dense points.  Sink particles
(Jappsen et al.\ 2005) are used to avoid the need for prohibitely high
resolutions while allowing for the simulation of individual or
clustered star formation.

The simulations start with a converging flow immersed in a much larger
box, so that the cloud can later interact freely with its environment,
without artificial confinement. The box is initially filled with WNM
gas at $n = 1 \pcc$ and $T = 5000$ K. The inflows have finite
durations, in order to be able to follow the subsequent evolution of
the cloud .  Non-equilibrium chemistry was not implemented, but
densities and temperatures are reached that allow quick formation of
molecular hydrogen (see Glover \& Mac Low 2007b).

\begin{figure}
\includegraphics[width=9cm,angle=0]{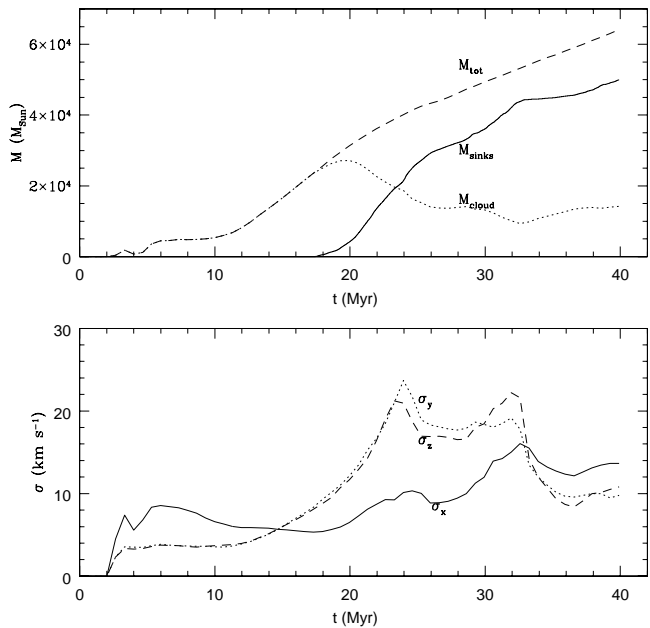}
\caption{{\it Top panel:} Evolution of the dense ($n > 50 \pcc$, {\it
dotted line}) gas mass, the mass in sink particles ({\it solid line}),
and the sum of the two ({\it dashed line}) for a simulation of dense
cloud formation with self-gravity in a 256 pc cubic box. The colliding
inflows had a diameter of 64 pc and a length of 112 pc each. {\it Bottom
panel:} Evolution of the velocity dispersion along the direction of the
colliding inflows ($x$) and the two directions perpendicular to it ($y$
and $z$). (From \VS\ et al 2007.)}
\label{fig:mass_sigma_evo}
\end{figure}

\begin{figure}
\includegraphics[width=9cm,angle=0]{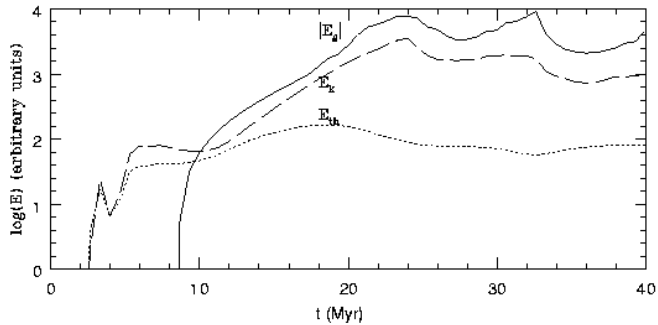}
\caption{Evolution of the thermal ({\it dotted line}), kinetic ({\it
dashed line}), and absolute value of the gravitational ({\it solid
line}) for the cloud formed in the same simulation as in Fig.\
\ref{fig:mass_sigma_evo}. (From \VS\ et al 2007.)}
\label{fig:energy_evo}
\end{figure}

The clouds formed in these simulations exhibit a secular evolutionary
process that proceeds along the following lines:

1. The turbulence in the dense clouds continues to be driven for as long
as the inflows last (see also Folini \& Walder 2006). The turbulent
velocity dispersion is maintained at a roughly constant level that
depends on the inflow Mach number, as can be seen for $0 \lesssim t
\lesssim 10$ Myr in the bottom panel of Fig. \ref{fig:mass_sigma_evo}.

2. The clouds {\it do not ever} reach an equilibrium state. As the gas
transits from the warm to the cold phase, it becomes denser and colder
than its surroundings, and the cloud's mass is constantly varying.
During the early stages, the mass increases continuously through
accretion from the inflows ($0 \le t \lesssim 20$ Myr in the top panel of
Fig.\ \ref{fig:mass_sigma_evo}). Later, the dense gas mass begins to
decrease because of its conversion to stars.

3. At some point during the evolution, the cloud's gravitational energy
becomes larger than the sum of its internal and turbulent energies; the
cloud becomes gravitationally unstable and begins to contract
($t\sim 10$ Myr in Fig.\ \ref{fig:energy_evo}).  This
contraction begins roughly at the time when the inflows have begun to
weaken, and the turbulence produced by the cloud formation
mechanism has begun to decay. This decay is however very slow, because
the flows weaken slowly, and so does the rate of turbulent energy
injection. The turbulence in the clouds is thus in a state intermediate between
being continuously driven and absolutely decaying. 

4. While the clouds are contracting, they exhibit a
near-equipartition energy balance satisfying $|E_{\rm g}| \approx 2
E_{\rm k}$ which {\it appears as} virial equilibrium. However, in this
case the energy balance is a signature of gravitational contraction
rather than of virial equilibrium, contrary to standard notions about
MCs (e.g., Blitz \& Williams 1999), and the velocity dispersion of the
cloud contains a dominant fraction of infall motion, rather than random
turbulence. 

5. Stars begin forming after the cloud has already been contracting a
long time (presumably still in atomic form). Specifically, the cloud
begins contracting at $t \sim 10$ Myr, while the first stars begin to
appear at $t \sim 17$ Myr. In this simulation, however, they form at
local density fluctuations produced by the initial turbulence (in turn
produced by the flow collision) long before the global collapse is
completed.

6. The star formation efficiency (SFE), that is, the mass converted
from gas into stars after some characteristic time, appears to be too
large, with 15\% of the mass having already been converted into stars
3 Myr after star formation starts ($t \sim 20$ Myr), and roughly three
times more mass in stars than in dense gas by $t \sim 40$
Myr. Instead, observational estimates of the SFE ranges from a few
percent for whole MC complexes (e.g., Myers et al.\ 1986), up to
30--50\% for cluster-forming cores (Lada \& Lada 2003). Thus, the
control of the SFE in the scenario of these simulations must probably
rely on magnetic support of the clouds (e.g., Mouschovias 1978; Shu et
al.\ 1987, Elmegreen 2007) or energy feedback from the stellar
products, as suggested by various authors (e.g., Norman \& Silk 1980;
Franco et al.\ 1994; Matzner \& McKee 2000; Hartmann et al.\ 2001;
Krumholz et al.\ 2006; Nakamura \& Li 2007).

\subsection{Formation of molecular hydrogen}

One argument often advanced in favour of cloud lifetimes longer than
10 Myr is the apparent difficulty involved in producing sufficient
H$_2$ in only 1--2~Myr to explain observed clouds, given the
relatively slow rate at which H$_2$ forms in the ISM.  The H$_2$
formation timescale in the ISM is approximately (Hollenbach, Werner,
\& Salpeter 1971)
\begin{equation}
 t_{\rm form} \simeq \frac{10^{9} \: {\rm yr}}{n},
\end{equation}
where $n$ is the number density in cm$^{-3}$, which suggests that in gas
with a mean number density $\bar{n} \sim 100 \: {\rm cm^{-3}}$, characteristic of 
most giant molecular clouds (Blitz \& Shu 1980), conversion from atomic to molecular form 
should take at least $10 \: {\rm Myr}$, longer than the entire lifetime of a transient 
cloud. However, estimates of this kind do not take account of dynamical processes such 
as supersonic turbulence or thermal instability.

\begin{figure}
\includegraphics[width=8cm,angle=270,scale=0.8]{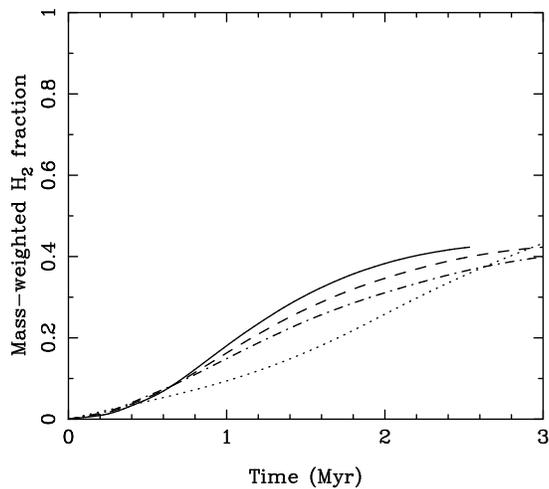}
\caption{Time evolution of the mass-weighted average H$_2$ fraction in
several runs with supersonically turbulent initial conditions from Glover \& Mac Low
(2007b; their Figure 1).  Runs using a local shielding approximation with 64$^3$ ({\em
dotted line}), 128$^3$ ({\em dash-dotted line}, 256$^3$ ({\em dashed
  line}), and 512$^3$ ({\em solid
line}) zones are shown.  All runs shown here use the local shielding
approximation, which works quite well for turbulent initial
conditions. Significant molecule formation occurs in 2~Myr, with some
regions having already become fully molecular.}
\label{turb-H2}
\end{figure}

Glover \& Mac Low (2007ab) used the ZEUS-MP MHD code (Norman 2000)
modified to include a simplified chemical network to follow the
non-equilibrium abundance of molecular hydrogen to study molecule
formation in a turbulent, self-gravitating flow with and without
magnetic fields.  They found that initially uniform gas does indeed
take tens of megayears to form molecules,  
but that initial supersonic turbulence produces density enhancements
that allow regions to become fully molecular within 3~Myr
(Fig.~\ref{turb-H2}), consistent with short timescales for molecular
cloud formation.

One crucial issue is the possibility that molecular clouds may
actually contain sizable amounts of atomic gas interspersed within the
molecular phase. Certainly, this is the impression that one gathers
from density fields like that shown in Fig. \ref{dens_sim}, where the
dense gas appears to have a fractal structure, with substructure
observed essentially down to the resolution limit.  Observations
appear to show atomic gas intermixed with the molecular gas
(e.g. Williams et al.\ 1995, Li \& Goldsmith 2003).  Hennebelle \&
Inutsuka (2006) have suggested that some of this gas may be warm, and
proposed that it could be heated by dissipation of MHD waves.
This issue remains open, but if confirmed, it
could have a deep impact on the dynamics and structure of molecular clouds.
Indeed, since the filling factor in molecular cloud is generally not 
greater than $\sim 10$\%, the density of the interclump medium  
must be smaller than a few times $\simeq 10$ cm$^{-3}$. Otherwise, the mass in the 
interclump medium would be comparable to or greater than the mass within the clumps 
(Williams et al. 1995 estimate that the interclump 
particle density is lower  than 10 cm$^{-3}$). 
Thus the assumption that the interclump medium has the same 
temperature than the gas inside the clumps, leads to a thermal pressure 
much lower than the thermal pressure of the standard ISM.

\subsection{Discussion and Implications} \label{sec:discussion}

The MC formation simulations discussed here are consistent with the
scenario of Hartmann et al.\ (2001): the formation of a GMC involves
accumulation of the gas from distances of up to hundreds of parsecs,
and takes up to 10--20 Myr to complete.  However, most of this process
occurs in the atomic phase, with molecular gas
only forming late in the evolution, when a sufficiently large column
density of dense gas has been collected (see also Franco \& Cox
1986). The formation of molecular gas 
occurs roughly simultaneously with the onset of star formation for
solar-neighborhood conditions.

If this scenario proves correct, it has two important implications.
First, the mechanism of cloud formation by accumulation and
condensation implies, in the presence of magnetic fields, that the
mass-to-flux ratio of the dense gas must {\it increase} with time.
This is consistent with the observation that diffuse CNM clouds are
generally magnetically subcritical (Heiles \& Troland 2005), while MCs
(which in this scenario begin their evolution as diffuse CNM clouds;
see \VS\ et al.\ 2006) are generally magnetically critical or
supercritical (Bourke et al.\ 2001; Crutcher 2004). Hartmann et al.\
(2001) suggest that the clouds in the solar neighborhood should become
supercritical roughly simultaneously with their becoming
molecular. The precise timing of the transition from subcritical to
supercritical, in relation to the time of becoming molecular and the
onset of star formation may be crucial in determining the
SFE. Moreover, the global cloud contraction observed in the
non-magnetic simulations may not take over until the clouds become
supercritical in the magnetic case.

Second, the domination of the evolution by global gravitational
contraction suggests a return to the scenario originally proposed by
Goldreich \& Kwan (1974), of global gravitational contraction for
MCs. This proposal was quickly dismissed by Zuckerman \& Palmer (1974)
through the argument that if all MCs converted all their mass into
stars in roughly one free-fall time, the resulting star formation rate
would be at least 10 times larger than that presently observed in the
Galaxy. Zuckerman \& Evans (1974) then proposed that the supersonic
linewidths in the clouds are produced primarily by local motions (the
hypothesis of microturbulence), which has been widely accepted until
recently.

However, several considerations argue against the microturbulent
picture. Turbulence is a regime of fluid flow characterized by having
the largest velocities at the driving scale, as indicated by the
negative slope of the turbulent energy spectrum both in the
incompressible (Kolmogorov 1941) and compressible (e.g., Passot \&
Pouquet 1987; Kritsuk et al.\ 2007) cases.  Comparisons of
observations with numerical simulations in various contexts show that
the dominant motions (and thus the driving scales) occur at scales the
size of the whole cloud or larger (e.g., Ossenkopf \& Mac Low 2002;
Heyer \& Brunt 2007).  This contradicts the microturbulent picture,
which requires small-scale driving.

Furthermore, evidence is beginning to accumulate that MCs or their
clumps may, in fact, be gravitationally collapsing. This has been
claimed for the Orion MC by Hartmann \& Burkert (2007), and for
NGC~2264 by Peretto et al.\ (2007).  If gravitational contraction
turns out to be a general feature of MCs, then the Goldreich \& Kwan
(1974; see also Field et al. 2006) suggestion may turn out to be
correct after all, at least for a subset of the clouds, or for the
star-forming regions of MCs (a large fraction of the volume of a MC is
devoid of star formation; e.g., Krumholz, Matzner, \& McKee 2006,
Elmegreen 2007). In this case, the regulation of the SFE may be
accomplished by the combined effects of magnetic support, of the
turbulence produced in the clouds during their formation, and of the
dispersive action of stellar feedback.

The initial turbulence in the cloud caused by its formation produces
nonlinear density fluctuations that can collapse before the global cloud
collapse is completed. The feedback from these first star formation
events may be able to suppress, or at least reduce, subsequent events
before all the gas of the cloud is turned into stars (Franco et al.\
1994). Moreover, in the presence of magnetic field fluctuations, parts
of the clouds may remain magnetically subcritical and thus supported
against collapse, while other parts may become supercritical as they
incorporate material from the surrounding WNM, and go into collapse. In
this case, the SFE in the locally supercritical regions may be large,
while the global average over whole GMCs may be low, because most of their
mass remains subcritical. The feedback from the active star forming
regions may then shred the clouds and leave subcritical fragments that
may collapse later, or even disperse away (Elmegreen 2007).

However, it is presently a matter of strong debate whether regulation
of the SFE by stellar energy input occurs by dispersal of the
star-forming clumps or by quasi-equilibrium support of the clouds. The
recent study by Nakamura \& Li (2007) suggests that near-equilibrium
can be achieved between driving by stellar outflows and the
self-gravity, but this study has a deep potential well and furthermore
uses periodic boundary conditions, so that the cloud cannot be
dispersed. Observationally, clusters older than several Myr are
generally observed to be devoid of gas, suggesting that they have been
able to disperse (or consume) their parent cloud (Leisawitz et al.\
1989; Hartmann et al.\ 2001).

These issues will hopefully be resolved in the near future
both through observations aimed at distinguishing these two scenarios,
and by numerical simulations of the entire evolution of MCs, from
formation to dispersal, including the feedback from stellar sources and
magnetic fields.



\end{document}